\documentclass[letterpaper,journal]{IEEEtran}

\usepackage{amsmath,amsfonts}
\usepackage{algorithm}
\usepackage{algorithmic}
\usepackage{array}
\usepackage{booktabs}
\usepackage{flushend}
\usepackage{stfloats}
\usepackage{cite}
\usepackage{graphicx}
\usepackage{needspace}
\usepackage[caption=false,font=normalsize,labelfont=sf,textfont=sf]{subfig}
\usepackage{textcomp}
\usepackage{threeparttable}
\usepackage{url}
\usepackage{collcell}
\usepackage{microtype}
\makeatletter
\renewcommand\section{\@startsection{section}{1}{\z@}%
  {2.0ex plus 0.4ex minus 0.4ex}%
  {0.5ex plus 0.2ex minus 0ex}%
  {\normalfont\normalsize\centering\scshape}}

\renewcommand\subsection{\@startsection{subsection}{2}{\z@}%
  {1.8ex plus 0.3ex minus 0.3ex}%
  {0.4ex plus 0.2ex minus 0ex}%
  {\normalfont\normalsize\itshape}}
\makeatother

\graphicspath{{figures/}}

\begin{document}
\flushbottom

\title{PVTSizing 2.0: A Fast and Reliable Analog Circuit Sizing Operator for Agentic EDA Workflows}

\author{

Zichen Kong\textsuperscript{1},
Minghuan Song\textsuperscript{2}, 
Xiyuan Tang\textsuperscript{3*},
Ziming Wei\textsuperscript{4}, 
Yi Zhong\textsuperscript{1},
Yuan Wang\textsuperscript{1*}
\\

\textsuperscript{1}School of Integrated Circuits, Peking University \\
\textsuperscript{2}Yuanpei College, Peking University\\
\textsuperscript{3}Institute for Artificial Intelligence, Peking University \\
\textsuperscript{4}School of Electronics Engineering and Computer Science, Peking University\\

\thanks{*Corresponding authors: \{xitang, wangyuan\}@pku.edu.cn}}


\maketitle

\begin{abstract}
Analog circuit design automation requires both coordinated tool execution and efficient numerical search. Large language models (LLMs) offer new ways to organize tool execution, but finding one sizing that satisfies specifications across process, voltage, and temperature (PVT) corners remains costly. PVTSizing 2.0 addresses these needs through a human--agent workflow built around an independently usable numerical optimizer. An agent carries the designer's approved task through execution and result review. The optimizer reduces unnecessary training through loss-guided updates and uses critic-guided local refinement to explore additional candidates while accounting for cross-corner conflicts. An optional LLM adviser supplies parameter relationships to focus numerical search on circuit-informed directions, while the optimizer selects adjustment amplitudes and verifies their effects through simulation. Across four circuit benchmarks, the standalone optimizer completes all 40 runs and achieves a $7.3\times$ speedup in overall mean runtime over the original PVTSizing, with a comparable mean sample count. In a separate study, the best tested knowledge-injection configuration further reduces mean runtime by 39.9\% and simulation samples by 35.4\% relative to the standalone optimizer.
\end{abstract}


\section{Introduction}
\begin{figure}[!t]
\centering
\includegraphics[width=\columnwidth]{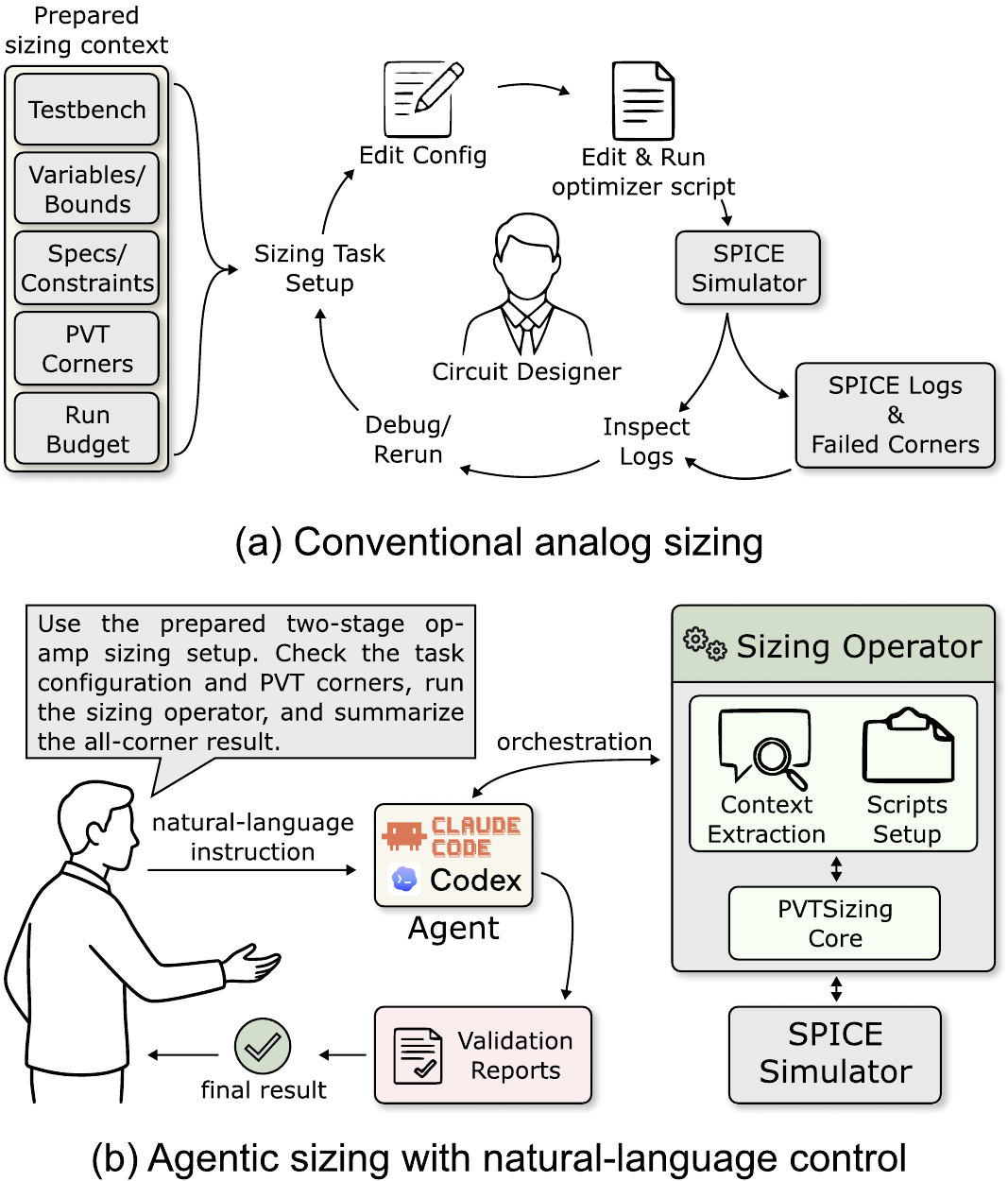}
\caption{Conventional and agent-assisted analog sizing workflows.}
\label{fig:motivation}
\end{figure}

Circuit design involves both solving design problems and organizing the information needed to solve them. A designer must keep circuit descriptions, specifications, and simulation results consistent as the design develops. Analog sizing makes this connection concrete: for a fixed topology, device parameters are adjusted until the simulated performance meets the specifications. Each adjustment must be translated into simulation inputs, and the resulting measurements must be collected for the next decision. Computer-aided design (CAD) tools have long supported this work. SPICE combines a machine-readable circuit description with numerical analysis~\cite{spice1973}; graphical interfaces, scripts, and tool application programming interfaces (APIs) make design data easier to prepare and exchange. Dedicated algorithms can make individual design decisions, yet designers still connect these capabilities into a working flow.

Large language models (LLMs) and agents offer a way to handle this coordination through existing tools: an agent can translate a design request into tool inputs, arrange execution, and return measurements for review. The numerical problems inside those tools still require effective search methods. Although LLM-based search has shown promise, studies of continuous and combinatorial optimization also report difficulties with increasing problem size and complex objective landscapes~\cite{opro}. A dedicated solver can carry out this search while the agent coordinates the surrounding task, as illustrated by the sizing flow in Fig.~\ref{fig:motivation}.

The cost of that solver then becomes central to the flow. In multi-PVT sizing, one set of device parameters must satisfy the specifications across process, voltage, and temperature (PVT) corners, even when an adjustment helps one corner but harms another. Numerical search learns from these measured responses; circuit understanding can complement it by suggesting which devices to adjust together. LLM circuit knowledge could therefore improve the search itself if the solver can translate it into adjustments and test their effects. This motivates three capabilities: human--agent collaboration to organize execution, an efficient numerical optimizer to solve the sizing problem, and a knowledge interface through which improved circuit reasoning can guide that optimizer.

Agentic electronic design automation (EDA) systems already connect design requests to tool execution. ChatEDA generates plans and scripts~\cite{chateda}, and LayoutCopilot supports interactive analog layout~\cite{layoutcopilot}. For sizing, simulation results must also inform the next parameter adjustment. Existing systems use this feedback to revise device parameters~\cite{eesizer} and combine circuit reasoning with numerical optimization~\cite{anaflow,astra}. Topology-aware collaboration~\cite{agenticsizing} and PVT checks in industrial tool flows~\cite{sable} bring circuit context and verification into this process. These developments establish a basis for coordinating sizing tasks; completing them efficiently also depends on how the underlying algorithm searches across corners.

Numerical sizing methods use measured performance to choose subsequent designs. Bayesian optimization (BO) selects candidates through surrogate models, while reinforcement learning (RL) learns parameter-selection policies~\cite{mace,autockt,gcnrl}. For multi-PVT sizing, measurements from different corners must guide the same design. RobustAnalog shares learning across corner tasks~\cite{robustanalog}, and RoSE-Opt combines circuit knowledge with full-condition feedback~\cite{roseopt}. Corner variation also enters policy training: AnaCraft trains sizing agents against a PVT agent~\cite{anacraft}, while specification priorities and graph representations guide learning across performance requirements and PVT conditions~\cite{prioritygraphrl}.

Runtime depends on the cost of both obtaining these measurements and learning from them. Batch acquisition and asynchronous execution improve simulation throughput~\cite{mace,apostle}, while PPAAS terminates unpromising evaluations early~\cite{ppaas}. The original PVTSizing reduces simulation work through pruning and combines trust-region BO (TuRBO) initialization with multi-task actor--critic search~\cite{pvtsizing_dac2024}. Within this framework, model training remains a source of computation: a fixed schedule continues the prescribed updates even when fitting errors become small. The same measurements describe how performance varies near explored designs, so they can also support local candidate searches alongside the actor. These additional candidates must be compared with the actor's proposals across corners, since a local improvement can create violations elsewhere.

LLMs can complement this measurement-based search with circuit reasoning, either by proposing candidates or by guiding the numerical solver. ADO-LLM combines LLM-generated candidates with BO proposals~\cite{ado_llm}; LLM-USO generates candidates from reusable circuit knowledge and ranks them numerically~\cite{llmuso}. Other methods let numerical optimization choose the sizes while the LLM guides the search: LEDRO adjusts search regions using calibration results and feedback~\cite{ledro}, and AutoSizer selects variables, parameter ranges, and inner optimization strategies~\cite{autosizer}. This separation allows circuit reasoning to guide numerical search without requiring the LLM to determine precise device sizes.

Circuit structure provides further guidance on how to organize that search. TopoSizing derives parameter relationships from topology and uses circuit knowledge in initialization and search-region adjustment~\cite{toposizing}. LASO-BOSS uses circuit understanding to plan stages with selected subcircuits, objectives, and budgets for numerical optimization~\cite{lasoboss}. In multi-PVT sizing, measured conflicts between corners give such guidance a concrete target: parameter relationships can suggest which devices to adjust together to address these conflicts. Expressing these suggestions as search directions leaves the numerical optimizer to determine adjustment amplitudes and test their cross-corner effects.

PVTSizing 2.0 brings task coordination and numerical search together in an agent-callable sizing operator. It extends the numerical core of PVTSizing and adds an optional knowledge interface, allowing the agent to invoke either standalone or knowledge-assisted optimization. The contributions are as follows:
\begin{itemize}
\item A \textbf{human--agent collaborative sizing workflow} reduces manual intervention by letting an agent prepare, launch, and monitor sizing tasks through a callable operator. Designers specify requirements, approve execution, and review measured results, while the agent handles routine tool coordination.
\item A \textbf{fast and reliable multi-PVT optimizer} handles sizing problems with over 40 design variables and a discrete search space containing more than $10^{100}$ parameter combinations. Loss-guided training reduces unnecessary updates, while critic-guided local refinement adds BO candidates and screens their cross-corner effects. A full-PVT test verifies successful results.
\item An \textbf{LLM-based knowledge-injection mechanism} provides a way to improve operator efficiency as LLM circuit reasoning advances. It translates circuit reasoning and measured conflicts into parameter directions, leaving adjustment amplitudes and simulation-based evaluation to numerical search. The optional interface accepts different advisers while preserving standalone operation.
\end{itemize}

The evaluation first examines the optimizer without knowledge injection, then the benefit of adding advice. Across four circuit benchmarks, the standalone optimizer completes all 40 runs and achieves a $7.3\times$ speedup in overall mean runtime over the original PVTSizing, with a comparable mean sample count. The best tested knowledge-injection configuration gives a further 39.9\% reduction in mean runtime and 35.4\% fewer simulation samples in a separate study. In both studies, success requires a sizing verified over the specified full PVT corner set. Section~\ref{sec:experiments} gives the recorded timing and execution conditions for these comparisons.

\begin{figure*}[!t]
\centering
\includegraphics[width=1\textwidth]{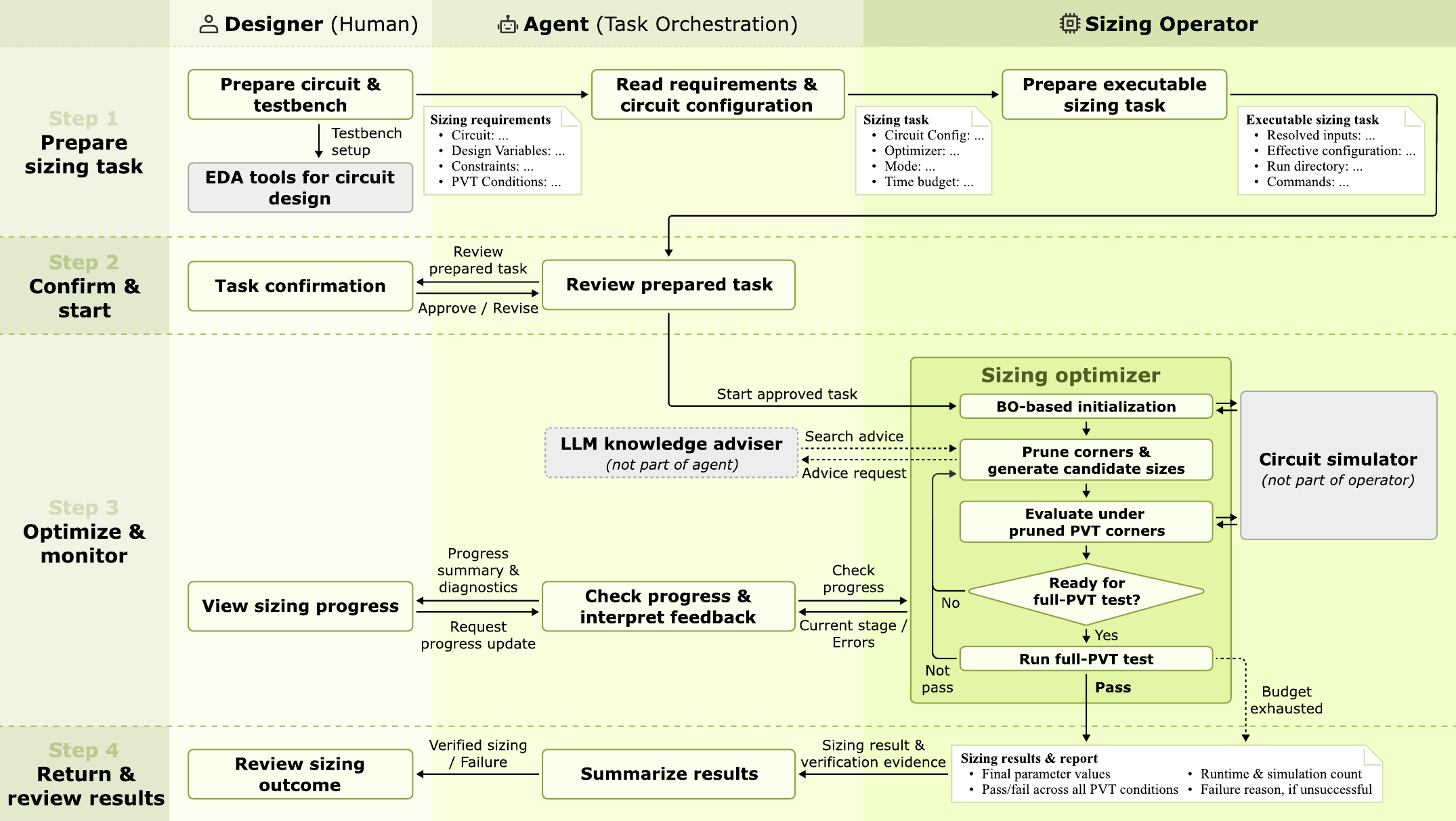}
\caption{Agentic sizing workflow and execution interface of PVTSizing 2.0.}
\label{fig:operator}
\end{figure*}

\begingroup
\setlength{\abovedisplayskip}{6pt plus 1pt minus 1pt}
\setlength{\belowdisplayskip}{6pt plus 1pt minus 1pt}
\setlength{\abovedisplayshortskip}{3pt plus 1pt}
\setlength{\belowdisplayshortskip}{5pt plus 1pt minus 1pt}
\section{Preliminaries}
\label{sec:background}

\subsection{Analog Sizing Formulation and Search Metrics}

A sizing that meets the specifications at one operating condition may fail at another. Multi-PVT sizing therefore seeks a single parameter vector that satisfies all required conditions for a fixed circuit topology. Parameter bounds and grids define the legal design space $D$. The simulation testbench evaluates a candidate $X=(X_1,\ldots,X_n)\in D$ over the specified PVT corner set $T=\{T_1,\ldots,T_K\}$, returning a performance vector $F(X\mid T_j)$ at each corner $T_j$.

The measured metrics can have different units and either lower or upper limits. To express their satisfaction or violation on a common scale, each specification is written as a scalar bound $C_i$ on a metric $F_i$, with a two-sided specification contributing two of the $m$ constraints. The normalized margin is defined as
\begin{equation}
\label{eq:normalized_margin}
h_{ij}(X)=
\begin{cases}
\dfrac{F_i(X\mid T_j)-C_i}{s_i(X\mid T_j)}, & i\in\mathcal{L},\\[6pt]
\dfrac{C_i-F_i(X\mid T_j)}{s_i(X\mid T_j)}, & i\in\mathcal{U},
\end{cases}
\end{equation}
where $\mathcal{L}$ and $\mathcal{U}$ index lower and upper bounds, respectively, and
\begin{equation}
s_i(X\mid T_j)=\max\!\left(|F_i(X\mid T_j)+C_i|,|C_i|,\epsilon\right).
\end{equation}
The scale uses $\epsilon>0$ to prevent a zero denominator. With this definition, a nonnegative margin indicates a satisfied constraint and a negative margin measures a violation, so the multi-PVT sizing problem becomes
\begin{equation}
\label{eq:multi_pvt_sizing}
\begin{aligned}
\textrm{find}\quad &X\in D\\
\textrm{s.t.}\quad &h_{ij}(X)\geq0,\quad i=1,\ldots,m,\ j=1,\ldots,K.
\end{aligned}
\end{equation}
A \emph{full-PVT test} checks this condition by evaluating one sizing over all of $T$. It passes only when every required measurement is valid and satisfies~\eqref{eq:multi_pvt_sizing}.

To guide the search toward satisfying these constraints, the native optimizer learns from a scalar figure of merit (FoM) at each corner:
\begin{equation}
\label{eq:condition_fom}
\mathrm{FoM}_j(X)=
\begin{cases}
f^+, & h_{ij}(X)\geq0,\ \forall i,\\
\displaystyle\sum_{i=1}^{m}\min\!\left(h_{ij}(X),0\right), & \textrm{otherwise}.
\end{cases}
\end{equation}
Negative scores accumulate the remaining violations, while $f^+=0.2$ marks a passing observation. This aggregate provides the learning signal; the original metric values remain available for constraint checks and diagnosis.

\subsection{Agentic EDA and Tool-Grounded Execution}

An LLM-based \emph{agent} combines a language model with task context and tools to carry out a sequence of actions. It reads the request and available observations, selects a tool operation, and uses the returned result to decide what to do next~\cite{react}. Retaining instructions and earlier results in the task context lets this loop continue across multiple operations and adapt to execution feedback.

Tools connect this reasoning loop to the design environment on a computer. File access exposes circuit configurations, netlists, logs, and simulation results; command-line tools and APIs allow scripts to edit design inputs, launch simulations, and query running jobs. A multimodal agent can also inspect screenshots and issue mouse or keyboard actions through a graphical interface~\cite{osworld}. The LLM chooses an operation and its arguments, while the software hosting the agent executes it through the corresponding interface and returns observations. For example, a simulation command can return a job identifier that the agent uses to query progress and retrieve measurements.

In this work, the agent coordinates a sizing task through a \emph{sizing operator}, which exposes task preparation, execution, status, and results through a callable interface. The numerical \emph{optimizer} inside the operator selects device parameters using simulation feedback. This separates task-level coordination from the repeated numerical decisions within a sizing run; Section~\ref{sec:operator_contract} describes their interaction in the workflow.

\endgroup

\section{Proposed PVTSizing 2.0}
\label{sec:pvtsizing2}

\subsection{Overview}

Fig.~\ref{fig:operator} shows how PVTSizing 2.0 carries a sizing task from the designer's requirements to a measured result. The agent handles routine preparation and execution through the operator, leaving the designer to confirm the task and review its outcome. Inside the operator, loss-guided training and critic-guided local refinement address training overhead and search progress, respectively. An optional LLM adviser further complements the standalone optimizer with circuit-informed search directions.

\subsection{Agentic Sizing Workflow}
\label{sec:operator_contract}

Much of the work surrounding numerical sizing consists of preparing text-based configurations, invoking tools, and collecting their outputs. Where these operations follow a defined procedure, an agent can carry them out through tool interfaces. PVTSizing 2.0 assigns this routine coordination to the agent so that the designer can focus on requirements and engineering judgment.

\emph{Step 1: Prepare the task.}
The designer supplies a circuit testbench, prepared in Cadence Virtuoso or another EDA tool, defining device models, analyses, and measurement expressions. The agent submits the sizing requirements and time budget to the operator, which reads the referenced circuit configuration and checks the inputs. It returns the prepared configuration, execution plan, and run identifier.

\emph{Step 2: Confirm and start.}
Through the operator, the agent reads the circuit information and effective sizing configuration, checks the variables and bounds, constraints, and PVT conditions against the requirements, and presents the task for confirmation. After incorporating any revisions, it launches the approved task through the operator.

\emph{Step 3: Optimize and monitor.}
Once launched, the operator runs the numerical search while the agent queries status and reads logs to report progress or execution failures. Within the search, the optimizer selects a sizing and its evaluation corners. Simulation scripts, such as OCEAN, apply the parameters and operating conditions, invoke the simulator, and export $F(X\mid T_j)$ for each evaluated corner. These measurements guide subsequent sizing decisions without requiring the agent to intervene in each numerical iteration.

\emph{Step 4: Return and review.}
At termination, the operator returns the sizing parameters and constraint results, together with the run status, runtime, and sample count. A successful run includes measurements from a passing full-PVT test; an unsuccessful run retains the observed violations. The agent brings these results back to the designer, who can assess the sizing or revise the task.

\subsection{Fast and Reliable Multi-PVT Optimization}
\label{sec:method}
Although the agent streamlines task preparation and execution, the underlying numerical search remains costly, even with the simulation pruning of the original PVTSizing~\cite{pvtsizing_dac2024}. PVTSizing 2.0 aims to shorten this time while maintaining the ability to find a feasible design. Repeated model training adds cost between simulations, and candidate quality determines whether search progresses toward feasibility. Loss-guided training avoids unnecessary updates, while critic-guided local refinement uses the same measurements to explore additional candidates and screen their effects across corners.

Algorithm~\ref{alg:pvtsizing} places these changes within the inherited search loop. Typical-corner BO supplies initial sizings, which are evaluated over $T$ to initialize the corner data. The actor then proposes sizings for individual corners, while the shared critic predicts their FoMs across corners.

For each corner $T_j$, a replay buffer $\mathcal{B}_j$ stores measurements for training, and a retained record $\ell_j=(X_{\mathrm{L}j},F_{\mathrm{L}j},\mathrm{FoM}_{\mathrm{L}j})$ holds one measured sizing and its performance to track progress. The records can refer to different sizings; a negative retained FoM identifies a failing corner. The simulation set $S\subseteq T$ contains the typical corner, all failing corners, and preceding-iteration sampling representatives whose records now pass. The loop retains K-means grouping for training-corner selection and critic-assisted actor pruning~\cite{pvtsizing_dac2024}; the following subsections explain the changes to model training and candidate selection.

\begin{algorithm}[!t]
\caption{Main numerical loop of PVTSizing 2.0}
\label{alg:pvtsizing}
\begin{algorithmic}[1]
\STATE \textbf{Input:} $D$, $T$, constraints, time budget, iteration threshold $E_{\max}$
\STATE Generate initial sizings at the typical corner using TuRBO
\STATE Evaluate them over $T$; initialize $\{\mathcal{B}_j,\ell_j\}_{j=1}^{K}$ and shared models
\STATE Train the models; set $e\gets0$
\WHILE{time remains}
    \STATE $e\gets e+1$; select active corners $S$ and training tasks
    \STATE Generate actor candidates and apply pruning
    \IF{no candidate passes pruning}
        \STATE Retain model weights; update on available training tasks; \textbf{continue}
    \ENDIF
    \STATE Generate local candidates when sufficient data exist
    \STATE Use critic-guided comparison to select $X$
    \STATE Map $X$ to legal parameter values; simulate over $S$
    \STATE Update corner records, replay buffers, and enabled local models
    \IF{all retained corner records are feasible or $e\geq E_{\max}$}
        \STATE Run a full-PVT test on $X$
        \IF{the test passes}
            \STATE \textbf{return} verified sizing $X$
        \ENDIF
        \IF{$e\geq E_{\max}$}
            \STATE \textbf{break}
        \ENDIF
        \STATE Incorporate the test measurements into corner records and replay buffers
    \ENDIF
    \STATE Apply loss-guided training to the current training tasks
\ENDWHILE
\STATE \textbf{return} no verified solution
\end{algorithmic}
\end{algorithm}

\subsubsection{Loss-Guided Model Training}

A fixed training schedule can continue updating models after their fitting errors have become small. Loss-guided training uses these errors to decide both how to allocate updates and when to stop. Since the actor learns from the critic, early rounds favor critic fitting; once its error is small, more updates go to the actor.

Each selected corner contributes a task: the critic fits its measured FoMs, and the actor follows the critic toward the feasible target $f^+$ in~\eqref{eq:condition_fom}. Both objectives use mean squared error, with PCGrad~\cite{pcgrad} combining task gradients as in the original method. A training call begins with three critic steps and one actor step per round, drawing a new replay minibatch at each step. Let $\overline{\mathcal{L}}_{Q,r}$ and $\overline{\mathcal{L}}_{A,r}$ denote the task-averaged losses at the last critic and actor steps of round $r$. Once $\overline{\mathcal{L}}_{Q,r}<0.1$, subsequent rounds in that call use one critic step and three actor steps.

A round counts toward stopping when both losses are below 0.02:
\begin{equation}
c_r=\sum_{q=1}^{r}\mathbf{1}\!\left[
\overline{\mathcal{L}}_{Q,q}<0.02\;\land\;
\overline{\mathcal{L}}_{A,q}<0.02\right].
\label{eq:training_stop}
\end{equation}
The call ends when $c_r\geq5$ or after 100 rounds, whichever occurs first; the qualifying rounds need not be consecutive.

\begin{figure*}[!t]
\centering
\includegraphics[width=\textwidth]{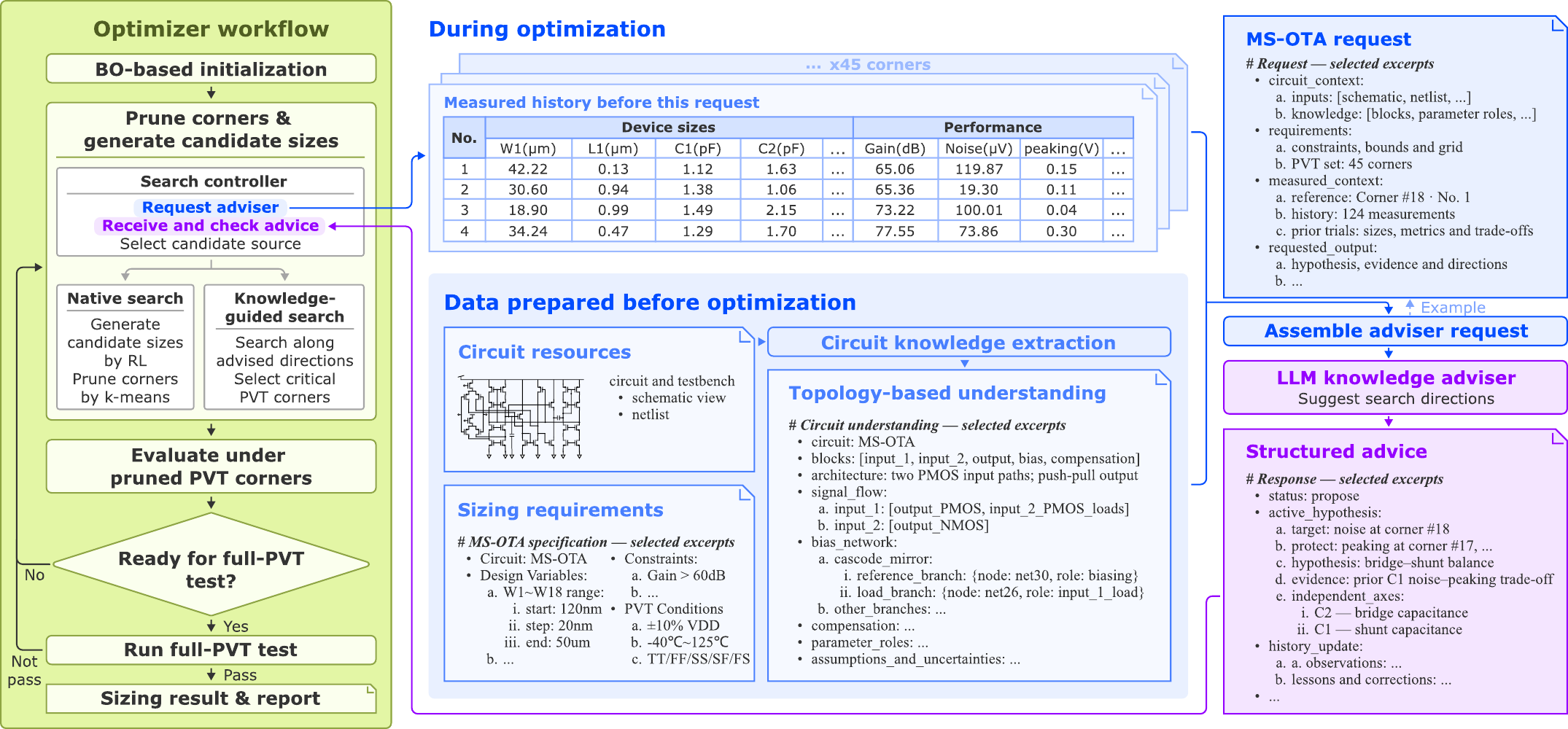}
\caption{LLM-based knowledge injection. Circuit context and measured history inform parameter directions; numerical search selects their amplitudes and compares candidates through simulation. The multi-stage operational transconductance amplifier (MS-OTA) panels show an illustrative request and response.}
\label{fig:knowledge_injection}
\end{figure*}

\subsubsection{Critic-Guided Local Refinement}
\label{sec:local_refinement}
While loss-guided training targets model-update overhead, search progress also depends on candidate quality. The shared actor learns a sizing policy from measurements across corners, but the same measurements can also support local BO around explored sizings. After actor pruning, per-corner TuRBO models~\cite{turbo} supply additional candidates. Before spending simulations on them, the critic screens their cross-corner effects, since improving one corner may cause violations elsewhere.

For candidate generation, one corner is sampled uniformly from $S$. Its TuRBO model is initialized from that corner's replay buffer on first use and supplies four candidates when at least 32 observations are available. Without local candidates, search uses the actor proposal selected by pruning.

The basic comparison favors candidates predicted to improve more failing corners. Let $\widehat f_j(X)$ be the critic's predicted FoM at $T_j$. Relative to the retained measurements, the predicted improvement count is
\begin{equation}
g(X)=\sum_{T_j\in S}\mathbf{1}\!\left[
\mathrm{FoM}_{\mathrm{L}j}<0\;\land\;
\widehat f_j(X)>\mathrm{FoM}_{\mathrm{L}j}\right].
\label{eq:local_candidate_gain}
\end{equation}
The local candidates are compared with the actor proposal for the configured typical corner. The candidate with the largest $g(X)$ is selected, retaining the actor on a tie. This score counts predicted improvements, not their magnitude or verified feasibility.

A candidate can improve failing corners yet lose feasibility elsewhere. A protected set $P$ therefore tracks recent failures, and $r_P(X)$ counts corners in $P$ whose retained records pass but whose predicted FoMs are negative. Ranking by $g(X)-r_P(X)$ accounts for these regressions alongside improvements. The set contains up to four corners that failed within the preceding six iterations. Protection is enabled when at most four corners currently fail and $|P|\geq2$, adding $P$ to $S$. Ties favor more predicted improvements, then a higher minimum predicted FoM over $P$.

The selected candidate is decoded, clipped to the bounds, and mapped onto the parameter grid for simulation over $S$. Its measured effects update the replay buffers, retained records, and enabled local models for the next iteration.

\subsubsection{Full-PVT Test and Termination}

Selective evaluation and single-corner search can leave different sizings in the corner records. Thus, every record can be feasible without one sizing passing all corners. When all records become feasible, a full-PVT test checks the current candidate $X$ over $T$. A passing test returns $X$; otherwise, its measurements guide further search.

After candidate evaluation, reaching the iteration threshold $E_{\max}$ also triggers this test, after which the optimizer stops.

\subsection{LLM-Based Knowledge Injection}
\label{sec:knowledge_injection}
To extend the optimizer beyond conventional numerical search, PVTSizing 2.0 can also leverage LLM-based circuit reasoning. Through the optional interface in Fig.~\ref{fig:knowledge_injection}, an adviser derives parameter directions from its interpretation of the circuit and measured conflicts. Numerical search then selects their amplitudes and tests the resulting sizings.
\subsubsection{Circuit Context and Adviser Requests}

Useful advice must connect the search's current difficulty to the circuit that produced it. Before optimization, circuit blocks and parameter roles are prepared through TopoSizing-style interpretation~\cite{toposizing} and checked against the schematic, parameterized netlist, and device-to-variable mapping. The operator reuses this context and its supporting resources with the specifications, parameter bounds and grids, and PVT conditions in later requests.

During search, stalled progress or conflicting corner responses can prompt a request. The optimizer selects a measured reference sizing $X_a$, favoring candidates with few remaining failed corners, and fixes a comparison set $G\subseteq T$ for the resulting trials. Up to four corners cover outstanding violations and conditions to protect from regression. Using this common set makes the advice trials comparable, while the native loop continues to use $S$ for its own simulations.

The request combines the reference measurements with the circuit context and the observed outcomes of previous advice. Calls are separated by new measurements and run asynchronously, allowing native search to continue while the adviser responds. A response superseded by progress in a full-PVT test is discarded.

\subsubsection{From LLM Advice to a Search Subspace}

Based on the circuit context and measured corner conflicts, the adviser identifies metrics to improve or protect and proposes one or two linearly independent parameter directions. Each direction specifies which parameters participate and their relative changes; the adviser can defer when the evidence does not support a useful hypothesis.

The MS-OTA example in Fig.~\ref{fig:knowledge_injection} illustrates how circuit reasoning supplies such directions. Increasing C1 reduced noise at one corner but worsened peaking at another. C1 shunts an internal node to ground, whereas C2 bridges the output and a mirror-reference node. These distinct connections suggest testing C2 alongside C1 to seek a better tradeoff. The advice proposes independent C1 and C2 directions; numerical search determines how much to change each while keeping other parameters fixed.

The optimizer checks that the advice matches its reference sizing and measurements and contains valid parameter names and finite, linearly independent directions. Their normalized columns form $B\in\mathbb{R}^{n\times d}$, where $d\in\{1,2\}$ and each column has maximum absolute coefficient one. Here, $B$ fixes the parameter relationships, while signed amplitudes $u$ determine their relative changes in log-parameter space. For positive-valued parameters,
\begin{equation}
X(u)=Q_D\!\left(X_a\odot\exp(Bu)\right),\qquad u\in\mathbb{R}^{d},
\label{eq:knowledge_subspace}
\end{equation}
where $\odot$ and $\exp$ act elementwise. Parameters absent from the directions stay at their values in $X_a$. Continuous proposals outside the original bounds are rejected, and $Q_D$ maps the remaining proposals onto the original parameter grid.

\subsubsection{Numerical Search and Feedback}

The proposed directions define where to search; measurements determine whether they are worth pursuing. Accepted advice starts a bounded cycle around the fixed reference $X_a$, temporarily taking over candidate evaluation from the native loop. Numerical trials select amplitudes $u$ and evaluate the resulting sizings on $G$, retaining improvements for a full-PVT test.

To compare these trials, the optimizer measures both the severity of violations and the number of failing corners. For valid measurements and nonzero bounds, the normalized violation $v_{ij}(X)$ is zero when constraint $i$ is met; otherwise, it is the measurement's absolute deviation from $C_i$ divided by $|C_i|$. The score combines the largest violation with the average:
\begin{equation}
J_G(X)=\max_{\substack{1\leq i\leq m\\T_j\in G}}v_{ij}(X)
+\frac{1}{m|G|}\sum_{T_j\in G}\sum_{i=1}^{m}v_{ij}(X).
\label{eq:knowledge_score}
\end{equation}
Lower $J_G$ indicates smaller violations on $G$, while the failed-corner count tracks progress toward feasibility. Nonfinite inputs are rejected, and a zero marking an unavailable measurement incurs an additional unit penalty.

Trial measurements guide subsequent amplitude choices. After at least three observations, a Gaussian-process model selects amplitudes using a lower-confidence-bound criterion. To compare the advised subspace with moves outside it, native alternatives are also evaluated on $G$ every fourth round, counted across cycles.

Only trials that improve $J_G$ or reduce the failed-corner count relative to $X_a$ are retained for full-PVT testing. Candidates are normally ranked first by $J_G$, then by failed-corner count. When $X_a$ is the best fully tested sizing by failed-corner count and has at most four failures, the priorities reverse to favor clearing the remaining corners.

The screening budget is four rounds and 32 new sizing--corner simulations per cycle, including diagnostics and missing reference measurements. Screening stops earlier if neither of the first two subspace trials improves on $X_a$. When a trial passes $G$ or screening ends, the best retained candidate, if any, is selected for the cycle's full-PVT test. This test reuses measurements for the same sizing and setup and simulates missing corners. A pass ends the run. Otherwise, the cycle ends when screening finishes or the test makes progress; until then, trials continue.

If no verified solution is found, native search resumes after the cycle. Trial measurements update its replay buffers, while the tested changes and their outcomes inform later adviser requests. Directions and observations remain reusable while the reference is current.

\section{Experimental Results}
\label{sec:experiments}
\begin{figure*}[!t]
\centering
\includegraphics[width=\textwidth]{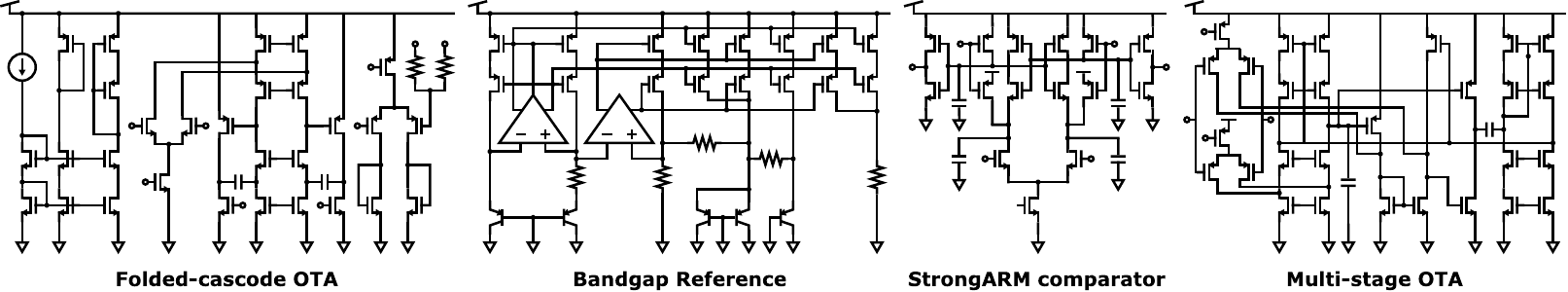}
\caption{Schematics of the four benchmark circuits}
\label{fig:benchmark_circuits}
\end{figure*}

\begin{table*}[!t]
\caption{Benchmark configurations and design specifications.}
\label{tab:benchmarks}
\centering
\begin{threeparttable}
\fontsize{8}{10}\selectfont
\setlength{\tabcolsep}{3pt}
\renewcommand{\arraystretch}{1.28}
\begin{tabular*}{\textwidth}{@{\extracolsep{\fill}}l c c c c c c c@{}}
\toprule
\multicolumn{8}{c}{\textit{(a) Benchmark configurations}} \\
\midrule
\textbf{Circuit}
& \textbf{Variable composition}\textsuperscript{a}
& \textbf{Search space size}
& \textbf{\#Cons.}\textsuperscript{b}
& \textbf{Process corners}
& $\boldsymbol{V_{\mathrm{DD}}}$ \textbf{(V)}
& \textbf{Temperature} ($^{\circ}$C)
& \textbf{\#Cond.} \\
\midrule
S55 FC-OTA  & $7W+7L+4M+C+R$ (20)  & $2.17\times10^{50}$  & 6  & TT/FF/SS/FS/SF & 1.08/1.20/1.32 & $-40/27/125$ & 45 \\
S55 BGR     & $10W+10L+5R$ (25)    & $2.06\times10^{68}$  & 4  & TT/FF/SS/FS/SF & 2.25/2.50/2.75 & $27$\textsuperscript{c} & 15 \\
S55 CMP     & $6W+6L+2C$ (14)      & $3.18\times10^{39}$  & 4  & TT/FF/SS/FS/SF & 1.08/1.20/1.32 & $-40/27/125$ & 45 \\
S55 MS-OTA  & $18W+18L+4M+2C$ (42) & $1.29\times10^{112}$ & 11 & TT/FF/SS/FS/SF & 1.08/1.20/1.32 & $-40/27/125$ & 45 \\
S40 FC-OTA  & $7W+7L+4M+C+R$ (20)  & $9.00\times10^{48}$  & 6  & TT/FF/SS/FS/SF & 0.99/1.10/1.21 & $-40/27/125$ & 45 \\
S130 FC-OTA & $7W+7L+4M+C+R$ (20)  & $1.58\times10^{49}$  & 6  & TT/FF/SS/FS/SF & 1.08/1.20/1.32 & $-40/27/125$ & 45 \\
S180 FC-OTA & $7W+7L+4M+C+R$ (20)  & $2.59\times10^{48}$  & 6  & TT/FF/SS/FS/SF & 1.62/1.80/1.98 & $-40/27/125$ & 45 \\
\bottomrule
\end{tabular*}

\vspace{3pt}
\begin{tabular*}{\textwidth}{@{\extracolsep{\fill}}l l@{}}
\multicolumn{2}{c}{\textit{(b) Design specifications}} \\
\midrule
\begin{minipage}[t]{0.485\textwidth}
\edef\specmetricwidth{\the\dimexpr\linewidth-49pt\relax}
\begin{tabular}[t]{@{}m{44pt}@{\hspace{5pt}}c@{}}
\textbf{Circuit} &
\begin{tabular}{@{}p{\dimexpr\specmetricwidth-64pt\relax}@{\hspace{6pt}}>{\centering\arraybackslash}p{58pt}@{}}
\textbf{Metric} & \textbf{Constraint}
\end{tabular} \\
\midrule
\shortstack[l]{FC-OTA\\(all nodes)} &
\begin{tabular}{@{}p{\dimexpr\specmetricwidth-64pt\relax}@{\hspace{6pt}}>{\centering\arraybackslash}p{58pt}@{}}
Power consumption & $\leq1~\mathrm{mW}$ \\
DC gain & $\geq70~\mathrm{dB}$ \\
Unity-gain bandwidth & $\geq20~\mathrm{MHz}$ \\
Output-referred noise & $\leq1~\mathrm{mV}$ \\
Phase margin & $60^{\circ}$--$90^{\circ}$
\end{tabular} \\
\cmidrule(lr){1-2}
BGR &
\begin{tabular}{@{}p{\dimexpr\specmetricwidth-64pt\relax}@{\hspace{6pt}}>{\centering\arraybackslash}p{58pt}@{}}
Temperature coefficient (TC) & $\leq25~\mathrm{ppm}/^{\circ}\mathrm{C}$ \\
Power-supply rejection ratio & $\geq10~\mathrm{dB}$ \\
Average output voltage & $0.5$--$2~\mathrm{V}$
\end{tabular} \\
\cmidrule(lr){1-2}
CMP &
\begin{tabular}{@{}p{\dimexpr\specmetricwidth-64pt\relax}@{\hspace{6pt}}>{\centering\arraybackslash}p{58pt}@{}}
Power consumption & $\leq2~\mu\mathrm{W}$ \\
Input-referred noise & $\leq100~\mu\mathrm{V}$ \\
Settling time & $\leq2~\mathrm{ns}$ \\
Propagation delay & $\leq2~\mathrm{ns}$
\end{tabular}
\end{tabular}
\end{minipage}
&
\begin{minipage}[t]{0.485\textwidth}
\edef\specmetricwidth{\the\dimexpr\linewidth-49pt\relax}
\begin{tabular}[t]{@{}m{44pt}@{\hspace{5pt}}c@{}}
\textbf{Circuit} &
\begin{tabular}{@{}p{\dimexpr\specmetricwidth-64pt\relax}@{\hspace{6pt}}>{\centering\arraybackslash}p{58pt}@{}}
\textbf{Metric} & \textbf{Constraint}
\end{tabular} \\
\midrule
MS-OTA &
\begin{tabular}{@{}p{\dimexpr\specmetricwidth-64pt\relax}@{\hspace{6pt}}>{\centering\arraybackslash}p{58pt}@{}}
Power consumption & $\leq400~\mu\mathrm{W}$ \\
DC gain & $\geq60~\mathrm{dB}$ \\
Unity-gain bandwidth & $\geq200~\mathrm{kHz}$ \\
Output-referred noise & $\leq300~\mu\mathrm{V}$ \\
Phase margin & $\geq50^{\circ}$ \\
Positive slew rate & $\geq0.5~\mathrm{V}/\mu\mathrm{s}$ \\
Negative slew rate & $\geq0.5~\mathrm{V}/\mu\mathrm{s}$ \\
Positive settling time & $\leq3~\mu\mathrm{s}$ \\
Negative settling time & $\leq3~\mu\mathrm{s}$ \\
Compensation capacitance & $\leq4~\mathrm{pF}$ \\
Normalized transient peaking & $\leq0.2$
\end{tabular}
\end{tabular}
\end{minipage} \\
\bottomrule
\end{tabular*}
\begin{tablenotes}[flushleft]
\fontsize{7}{8.6}\selectfont
\item[\textsuperscript{a}] $W$, $L$, $M$, $C$, and $R$ denote transistor widths, lengths, multiplicities, capacitors, and resistors, respectively.
\item[\textsuperscript{b}] Two-sided ranges count as two scalar constraints.
\item[\textsuperscript{c}] BGR varies process and voltage across 15 conditions. Temperature is swept internally from $-40$ to $125\,^{\circ}$C to measure TC; it is not an additional corner dimension.
\end{tablenotes}
\end{threeparttable}
\end{table*}

\subsection{Experimental Setup}

To evaluate the effectiveness and efficiency of PVTSizing 2.0 in multi-PVT sizing, experiments were conducted on four real-world analog circuits: a folded-cascode operational transconductance amplifier (FC-OTA), a bandgap reference (BGR), a StrongARM comparator (CMP), and a multi-stage OTA (MS-OTA). These circuits cover AC performance, reference-voltage stability, clocked timing requirements, and transient response. Their sizing tasks range from 14 to 42 design variables and from four to eleven scalar constraints. The largest discrete design space contains more than $10^{100}$ parameter combinations.

The four-circuit comparison uses SMIC55 to evaluate different topologies under a common technology. FC-OTA is also sized independently in SMIC40, SMIC130, and SMIC180 to examine different process nodes. The PVT configurations combine five process corners (TT/FF/SS/FS/SF) with three supply voltages and, for FC-OTA, CMP, and MS-OTA, three temperatures, giving 45 conditions. BGR instead uses 15 process–voltage conditions, each with an internal temperature sweep for TC measurement. The SMIC55 circuits are shown in Fig.~\ref{fig:benchmark_circuits}, and Tab.~\ref{tab:benchmarks} summarizes the design variables, PVT configurations, and performance specifications of the benchmark circuits.

The experiments were conducted on a server running CentOS 7.9, equipped with two 16-core Intel Xeon Gold 6246R processors at 3.40~GHz and 256~GiB of memory. Circuit testbenches were prepared in Cadence Virtuoso and simulated with Spectre, using up to eight concurrent simulation workers per run.

The standalone PVTSizing 2.0 optimizer is compared with the original PVTSizing~\cite{pvtsizing_dac2024}, RobustAnalog~\cite{robustanalog}, and five optimizers available in Virtuoso~\cite{virtuoso_optimization}. Each method is evaluated over ten independent runs per circuit, with a 10-h time budget and an iteration threshold of 1000. Knowledge injection is disabled in the main comparison and component ablations.

The PVTSizing 2.0 actor and critic each have four hidden layers of 500, 600, 600, and 200 units. Both use Adam with a learning rate of $3\times10^{-4}$. A run is considered successful only if its sizing passes the full-PVT test. Success is reported as successful/total runs, while sample counts and runtime are averaged over successful runs only. One sample corresponds to evaluating a sizing at one PVT condition, and sample counts include initialization, multi-PVT search, and full-PVT tests.

\begin{table*}[!t]
\caption{Sizing results on four real-world circuits\textsuperscript{1}}
\label{tab:s55_main_baseline}
\centering
\begin{threeparttable}
\fontsize{8}{10}\selectfont
\renewcommand{\arraystretch}{1.18}
\setlength{\tabcolsep}{0pt}

\newcommand{\igap}{\hspace{1pt}}
\newcommand{\ggap}{\hspace{5pt}}

\newcommand{\succcell}[1]{\makebox[\linewidth][c]{\makebox[18pt][r]{#1}}}
\newcommand{\sampcell}[1]{\makebox[\linewidth][c]{\makebox[29pt][r]{#1}}}
\newcommand{\timecell}[1]{\makebox[\linewidth][c]{\makebox[24pt][r]{#1}}}
\newcolumntype{U}{>{\collectcell\succcell}p{20pt}<{\endcollectcell}}
\newcolumntype{V}{>{\collectcell\sampcell}p{29pt}<{\endcollectcell}}
\newcolumntype{Z}{>{\collectcell\timecell}p{26pt}<{\endcollectcell}}

\begin{tabular}{
@{}>{\raggedright\arraybackslash}p{95pt}@{\hspace{3.5pt}}
*{4}{U@{\igap}V@{\igap}Z@{\ggap}}
U@{\igap}V@{\igap}Z@{}
}
\toprule
\textbf{Method} &
\multicolumn{3}{c}{\textbf{FC-OTA}} &
\multicolumn{3}{c}{\textbf{BGR}} &
\multicolumn{3}{c}{\textbf{CMP}} &
\multicolumn{3}{c}{\textbf{MS-OTA}} &
\multicolumn{3}{c}{\textbf{Overall}\tnote{2}} \\
\cmidrule(l{3pt}r{3pt}){2-4}\cmidrule(l{3pt}r{3pt}){5-7}\cmidrule(l{3pt}r{3pt}){8-10}\cmidrule(l{3pt}r{3pt}){11-13}\cmidrule(l{3pt}r{3pt}){14-16}
& \multicolumn{1}{c}{\fontsize{7}{9}\selectfont Succ.\,$\uparrow$}
& \multicolumn{1}{c}{\fontsize{7}{9}\selectfont Samp.\,$\downarrow$}
& \multicolumn{1}{c@{\hspace{4pt}}}{\fontsize{7}{9}\selectfont Time (h)$\downarrow$}
& \multicolumn{1}{c}{\fontsize{7}{9}\selectfont Succ.\,$\uparrow$}
& \multicolumn{1}{c}{\fontsize{7}{9}\selectfont Samp.\,$\downarrow$}
& \multicolumn{1}{c@{\hspace{4pt}}}{\fontsize{7}{9}\selectfont Time (h)$\downarrow$}
& \multicolumn{1}{c}{\fontsize{7}{9}\selectfont Succ.\,$\uparrow$}
& \multicolumn{1}{c}{\fontsize{7}{9}\selectfont Samp.\,$\downarrow$}
& \multicolumn{1}{c@{\hspace{4pt}}}{\fontsize{7}{9}\selectfont Time (h)$\downarrow$}
& \multicolumn{1}{c}{\fontsize{7}{9}\selectfont Succ.\,$\uparrow$}
& \multicolumn{1}{c}{\fontsize{7}{9}\selectfont Samp.\,$\downarrow$}
& \multicolumn{1}{c@{\hspace{4pt}}}{\fontsize{7}{9}\selectfont Time (h)$\downarrow$}
& \multicolumn{1}{c}{\fontsize{7}{9}\selectfont Succ.\,$\uparrow$}
& \multicolumn{1}{c}{\fontsize{7}{9}\selectfont Samp.\,$\downarrow$}
& \multicolumn{1}{c}{\fontsize{7}{9}\selectfont Time (h)$\downarrow$} \\
\addlinespace[1pt]
\midrule
Virtuoso\textsuperscript{3}: Global & 8/10 & 3,862.6 & 0.570 & 1/10 & 14,670.0 & 5.696 & 9/10 & 2,573.2 & 0.594 & 4/10 & 17,659.3 & 5.637 & 22/40 & 9,691.3 & 3.124 \\
Virtuoso: BFGS & \textbf{10/10} & 15,227.0 & 2.191 & 0/10 & \multicolumn{1}{c}{\textemdash\textsuperscript{4}} & \multicolumn{1}{c}{\textemdash} & \textbf{10/10} & 7,056.4 & 1.932 & 0/10 & \multicolumn{1}{c}{\textemdash} & \multicolumn{1}{c}{\textemdash} & 20/40 & \multicolumn{1}{c}{\textemdash} & \multicolumn{1}{c}{\textemdash} \\
Virtuoso: Hooke--Jeeves & \textbf{10/10} & 8,707.5 & 1.197 & 2/10 & 4,422.5 & 1.791 & 9/10 & 7,386.1 & 1.692 & 2/10 & 21,002.5 & 7.057 & 23/40 & 10,379.7 & 2.934 \\
Virtuoso: Conjugate gradient & \textbf{10/10} & 7,271.1 & 1.041 & 0/10 & \multicolumn{1}{c}{\textemdash} & \multicolumn{1}{c}{\textemdash} & \textbf{10/10} & 5,499.2 & 1.550 & 1/10 & 15,465.0 & 3.986 & 21/40 & \multicolumn{1}{c}{\textemdash} & \multicolumn{1}{c}{\textemdash} \\
Virtuoso: Brent--Powell & 9/10 & 5,618.1 & 1.364 & 0/10 & \multicolumn{1}{c}{\textemdash} & \multicolumn{1}{c}{\textemdash} & \textbf{10/10} & 2,355.5 & 1.225 & 2/10 & 22,383.0 & 8.312 & 21/40 & \multicolumn{1}{c}{\textemdash} & \multicolumn{1}{c}{\textemdash} \\
RobustAnalog & \textbf{10/10} & 3,783.6 & 0.471 & 7/10 & 11,150.3 & 4.048 & \textbf{10/10} & 2,278.0 & 0.654 & \textbf{10/10} & 9,030.1 & 3.518 & 37/40 & 6,560.5 & 2.173 \\
PVTSizing & \textbf{10/10} & \textbf{985.2} & 1.628 & \textbf{10/10} & 2,047.5 & 7.188 & \textbf{10/10} & 1,034.8 & 1.891 & 8/10 & \textbf{1,970.3} & 5.138 & 38/40 & \textbf{1,509.4} & 3.961 \\
\textbf{PVTSizing 2.0 (Proposed)} & \textbf{10/10} & 1,051.5 & \textbf{0.233} & \textbf{10/10} & \textbf{1,285.1} & \textbf{0.547} & \textbf{10/10} & \textbf{821.9} & \textbf{0.175} & \textbf{10/10} & 2,964.4 & \textbf{1.211} & \textbf{40/40} & 1,530.7 & \textbf{0.541} \\
\bottomrule
\end{tabular}
\vspace{3pt}
\renewcommand{\TPTnoteSettings}{%
  \setlength{\leftmargin}{7pt}%
  \setlength{\labelwidth}{5pt}%
  \setlength{\labelsep}{2pt}%
  \setlength{\itemindent}{0pt}%
  \setlength{\itemsep}{0pt}%
  \setlength{\parsep}{0pt}%
}
\begin{tablenotes}
\fontsize{7}{8.6}\selectfont
\item[1]Success reports successful/total runs (10 per method and circuit). Samples and time are means over successful runs only.
\item[2]Overall costs are the unweighted arithmetic means of the four circuit-level means.
\item[3]Virtuoso Size-over-Corners.
\item[4]An em dash indicates no successful run for a circuit, or an undefined Overall cost when a circuit-level mean is unavailable.
\end{tablenotes}
\end{threeparttable}
\end{table*}

\Needspace*{4\baselineskip}
\subsection{Multi-PVT Sizing Performance}

Tab.~\ref{tab:s55_main_baseline} presents the sizing results of different optimization methods across the four benchmark circuits. PVTSizing 2.0 completes all 40 runs, compared with 38 for the original PVTSizing, 37 for RobustAnalog, and 20–23 for the Virtuoso optimizers. BGR is highly sensitive to parameter changes, while MS-OTA presents a high-dimensional sizing problem with 42 design variables. These challenges are reflected in the lower completion rates of several competing methods on the two circuits. In contrast, PVTSizing 2.0 completes all ten runs on both circuits, demonstrating its reliability in challenging sizing tasks.

Beyond the success rates, Tab.~\ref{tab:s55_main_baseline} shows a substantial improvement in sizing efficiency. Compared with the original PVTSizing, PVTSizing 2.0 reduces the overall mean runtime from 3.961 to 0.541~h, achieving a $7.3\times$ speedup, while maintaining a comparable mean sample count (1,509.4 versus 1,530.7). This distinction is particularly evident in MS-OTA, where PVTSizing 2.0 completes sizing much faster despite using more samples. These results indicate that reducing simulation samples alone is insufficient to characterize optimization efficiency, as computational effort between simulations also contributes to runtime. Compared with RobustAnalog, PVTSizing 2.0 achieves both fewer samples and shorter runtime across all four circuits, reducing the overall mean sample count by 76.7\% and achieving a $4.0\times$ speedup.

Tab.~\ref{tab:cross_process} further evaluates FC-OTA sizing across four process nodes, with independent optimization performed using node-specific device models and operating conditions. All ten independent runs at each node pass the full-PVT test, with mean runtime ranging from 0.233 to 0.682~h. Although the required search effort varies across technologies, feasible sizings are consistently obtained, indicating that the optimizer's effectiveness is not limited to a particular process node.

\begin{table}[!t]
\caption{FC-OTA sizing results across process nodes}
\label{tab:cross_process}
\centering
\fontsize{8}{10}\selectfont
\renewcommand{\arraystretch}{1.18}
\setlength{\tabcolsep}{0pt}

\newcommand{\succcell}[1]{\makebox[\linewidth][c]{\makebox[18pt][r]{#1}}}
\newcommand{\sampcell}[1]{\makebox[\linewidth][c]{\makebox[29pt][r]{#1}}}
\newcommand{\timecell}[1]{\makebox[\linewidth][c]{\makebox[24pt][r]{#1}}}
\newcolumntype{U}{>{\collectcell\succcell}p{22pt}<{\endcollectcell}}
\newcolumntype{V}{>{\collectcell\sampcell}p{34pt}<{\endcollectcell}}
\newcolumntype{Z}{>{\collectcell\timecell}p{30pt}<{\endcollectcell}}

\begin{tabular}{@{}>{\raggedright\arraybackslash}p{58pt}@{\hspace{7pt}}U@{\hspace{5pt}}V@{\hspace{5pt}}Z@{}}
\toprule
\textbf{Process node} &
\multicolumn{1}{c}{\fontsize{7}{9}\selectfont Succ.\,$\uparrow$} &
\multicolumn{1}{c}{\fontsize{7}{9}\selectfont Samp.\,$\downarrow$} &
\multicolumn{1}{c}{\fontsize{7}{9}\selectfont Time (h)\,$\downarrow$} \\
\midrule
SMIC40  & 10/10 & 923.7   & 0.367 \\
SMIC55  & 10/10 & 1,051.5 & 0.233 \\
SMIC130 & 10/10 & 844.1 & 0.267 \\
SMIC180 & 10/10 & 1,370.0 & 0.682 \\
\bottomrule
\end{tabular}
\end{table}

\subsection{Ablation Study}
\label{sec:ablation}

The MS-OTA ablations examine how loss-guided training and critic-guided local refinement contribute to runtime and completion. The first variant replaces loss-guided training with the original fixed training schedule~\cite{pvtsizing_dac2024}. The second removes the per-corner local BO models and their candidate-generation path. Both variants retain the full optimizer's initialization procedure, pruning, specifications, and full-PVT test. Each starts ten independent runs from scratch under the 10-h limit, with the main-table MS-OTA results as the reference.

\begin{table}[!t]
\caption{Ablation of the optimizer components on MS-OTA}
\label{tab:ablation}
\centering
\begin{threeparttable}
\fontsize{8}{10}\selectfont
\renewcommand{\arraystretch}{1.18}
\setlength{\tabcolsep}{0pt}
\newcommand{\succcell}[1]{\makebox[\linewidth][c]{\makebox[18pt][r]{#1}}}
\newcommand{\sampcell}[1]{\makebox[\linewidth][c]{\makebox[29pt][r]{#1}}}
\newcommand{\timecell}[1]{\makebox[\linewidth][c]{\makebox[24pt][r]{#1}}}
\newcolumntype{U}{>{\collectcell\succcell}p{22pt}<{\endcollectcell}}
\newcolumntype{V}{>{\collectcell\sampcell}p{29pt}<{\endcollectcell}}
\newcolumntype{T}{>{\collectcell\timecell}p{34pt}<{\endcollectcell}}
\newcolumntype{E}{>{\collectcell\timecell}p{47pt}<{\endcollectcell}}

\begin{tabular}{@{}>{\raggedright\arraybackslash}p{112pt}
@{\hspace{1pt}}U@{\hspace{1pt}}V@{\hspace{1pt}}T@{\hspace{1pt}}E@{}}
\toprule
\textbf{Configuration} &
\multicolumn{1}{c}{\fontsize{7}{9}\selectfont Succ.\,$\uparrow$} &
\multicolumn{1}{c}{\fontsize{7}{9}\selectfont Samp.\,$\downarrow$} &
\multicolumn{1}{c}{\fontsize{7}{9}\selectfont Time (h)\,$\downarrow$} &
\multicolumn{1}{c}{\fontsize{7}{9}\selectfont Exp.\ time (h)\tnote{1}\,$\downarrow$} \\
\midrule
\textbf{Full method (PVTSizing 2.0)} & \textbf{10/10} & 2,964.4 & 1.211 & \textbf{1.211} \\
w/o loss-guided training & 9/10 & \textbf{1,823.1} & 3.644 & 4.280 \\
w/o critic-guided local refinement & 7/10 & 2,311.4 & \textbf{1.028} & 3.719 \\
\bottomrule
\end{tabular}
\vspace{3pt}
\renewcommand{\TPTnoteSettings}{%
  \setlength{\leftmargin}{7pt}%
  \setlength{\labelwidth}{5pt}%
  \setlength{\labelsep}{2pt}%
  \setlength{\itemindent}{0pt}%
  \setlength{\itemsep}{0pt}%
  \setlength{\parsep}{0pt}%
}
\begin{tablenotes}
\fontsize{7}{8.6}\selectfont
\item[1]Expected time is averaged over all ten runs, using the measured
runtime for successful runs and 10~h for unsuccessful runs.
\end{tablenotes}
\end{threeparttable}
\end{table}

In Tab.~\ref{tab:ablation}, restoring the fixed training schedule reduces samples among successful runs, but raises their mean runtime from 1.211 to 3.644~h. Nine runs succeed, and the remaining run reaches the time limit. Removing local refinement produces a different tradeoff. The successful-run mean falls to 1.028~h, but only seven runs succeed. The apparent speed advantage therefore applies only to the successful subset.

To compare runtime together with completion, expected time assigns 10~h to each failure and uses measured runtime for each success. Under this convention, the full optimizer has the lowest expected time, 1.211~h, compared with 4.280~h without loss-guided training and 3.719~h without local refinement.

\Needspace*{4\baselineskip}
\subsection{LLM-Based Knowledge Injection}
\label{sec:knowledge_results}

To investigate how LLM reasoning capabilities affect knowledge-assisted sizing, three models were selected to represent an expected progression in circuit-reasoning ability: GPT-5.5, GPT-5.6 Terra, and GPT-6 Astra. All three use low reasoning effort and provide advice through the same interface. BO initialization remains unchanged, and advice is introduced only during multi-PVT search.

\begin{table}[!t]
\caption{Knowledge injection on MS-OTA}
\label{tab:knowledge_hints}
\centering
\begin{threeparttable}
\fontsize{8}{10}\selectfont
\renewcommand{\arraystretch}{1.15}
\setlength{\tabcolsep}{0pt}

\newcommand{\succcell}[1]{\makebox[\linewidth][c]{\makebox[18pt][r]{#1}}}
\newcommand{\timecell}[1]{\makebox[\linewidth][c]{\makebox[23pt][r]{#1}}}
\newcommand{\sampcell}[1]{\makebox[\linewidth][c]{\makebox[29pt][r]{#1}}}
\newcommand{\itercell}[1]{\makebox[\linewidth][c]{\makebox[23pt][r]{#1}}}
\newcommand{\callcell}[1]{\makebox[\linewidth][c]{\makebox[20pt][r]{#1}}}

\newcolumntype{U}{>{\collectcell\succcell}p{21pt}<{\endcollectcell}}
\newcolumntype{T}{>{\collectcell\timecell}p{27pt}<{\endcollectcell}}
\newcolumntype{S}{>{\collectcell\sampcell}p{33pt}<{\endcollectcell}}
\newcolumntype{I}{>{\collectcell\itercell}p{26pt}<{\endcollectcell}}
\newcolumntype{C}{>{\collectcell\callcell}p{24pt}<{\endcollectcell}}

\begin{tabular}{
@{}>{\raggedright\arraybackslash}p{75pt}@{\hspace{5pt}}
U@{\hspace{4pt}}
T@{\hspace{3pt}}
S@{\hspace{3pt}}
I@{\hspace{5pt}}
C@{}
}

\toprule
\textbf{Configuration} &
\multicolumn{1}{c}{\fontsize{7}{9}\selectfont Succ.\,$\uparrow$} &
\multicolumn{1}{c}{\fontsize{7}{9}\selectfont Time (h)\,$\downarrow$} &
\multicolumn{1}{c}{\fontsize{7}{9}\selectfont Samp.\,$\downarrow$} &
\multicolumn{1}{c}{\fontsize{7}{9}\selectfont Iter.\,$\downarrow$} &
\multicolumn{1}{c}{\fontsize{7}{9}\selectfont \# API Calls\,$\downarrow$} \\
\midrule

PVTSizing 2.0 only &
10/10 & 1.211 & 2,964.4 & 250.4 & \textemdash \\
\addlinespace[1.5pt]
PVTSizing 2.0 + KI\tnote{1} & & & & & \\
\hspace{1em} \textit{w/ GPT-5.5} &
10/10 & 1.225 & 2,945.5 & 141.1 & 9.6 \\
\hspace{1em} \textit{w/ GPT-5.6 Terra} &
10/10 & 0.983 & 2,457.1 & 138.0 & 10.8 \\
\hspace{1em} \textit{w/ GPT-6 Astra} &
10/10 & \textbf{0.728} & \textbf{1,914.3} & \textbf{100.8} & \textbf{6.4} \\

\bottomrule
\end{tabular}

\vspace{2pt}
\begin{tablenotes}
\fontsize{7}{8.6}\selectfont
\item[1] KI denotes Knowledge Injection.
\end{tablenotes}
\end{threeparttable}
\end{table}

As shown in Tab.~\ref{tab:knowledge_hints}, two of the three knowledge-assisted configurations further reduce both runtime and simulation samples compared with the standalone optimizer. GPT-5.6 Terra reduces mean runtime by 18.8\% and samples by 17.1\%, while GPT-6 Astra achieves larger reductions of 39.9\% and 35.4\%, respectively. In contrast, GPT-5.5 produces nearly the same sample count as the standalone optimizer, with a slightly longer runtime of 1.225~h versus 1.211~h. Although all three configurations reduce the recorded number of PVT iterations, fewer iterations do not necessarily lead to lower search costs. These results suggest that LLM-based knowledge injection can further improve sizing efficiency, but its benefit depends on how effectively the suggested directions guide subsequent numerical search.

\section{Discussion}
\label{sec:discussion}

These results suggest an additional route for improving numerical sizing: an optimizer can benefit from better circuit advice as well as improvements to its own search algorithm. Through the advice interface, a model may identify parameter relationships that give the optimizer more useful directions to explore. The optimizer still selects the adjustments and tests the resulting designs, and it can run independently when advice is unavailable. This provides a way to improve the knowledge available to the search without redesigning the optimizer for each model. 

The time spent obtaining and testing advice must also be counted. A direction that saves simulations may require substantial numerical effort to find useful adjustments. Similar choices arise within the optimizer itself. Reducing training overhead can shorten execution, while spending more effort on candidate search can help more runs reach feasibility. Both can be worthwhile when they help complete the sizing task within the budget. Evaluating these choices requires runtime and completion alongside sample count, since fewer samples alone do not indicate a faster or more reliable search.

In practical use, the budget must cover simulation, surrogate-model training, and LLM inference. Cloud inference incurs service charges, often based on token usage; local inference requires hardware capacity and incurs operating costs. A more capable adviser may save search effort but take longer or cost more to produce its advice. More parallel computation can likewise shorten turnaround time while consuming additional resources.

The choice of model also affects whether a run can be repeated. Model updates or different responses to the same request can change later search decisions. Recording model configurations, requests, responses, and executed actions helps trace those decisions and replay recorded advice. Repeating the complete search also requires control over the numerical environment. 

For confidential designs, the deployment environment may be decided before a model is selected. Circuit and process information that cannot leave a controlled environment requires an adviser that can run within that boundary. This may restrict model choices and require additional local computing resources. 

To sum up, future design flows should allow the adviser, its deployment, and the extent of its involvement to be chosen according to the project’s needs for speed, cost, reproducibility, and data protection. Numerical verification remains the common requirement for accepting the resulting design.

\section{Conclusion}
\label{sec:conclusion}

PVTSizing 2.0 addresses task coordination and numerical sizing within a human--agent workflow. The agent carries the designer's approved task through execution, while an independently usable optimizer searches for a sizing verified across the required PVT corners. LLM circuit advice can extend that search through parameter relationships whose adjustments are selected and tested numerically. In the reported four-circuit comparison, the standalone optimizer completes all 40 runs with a $7.3\times$ speedup in overall mean runtime over the original PVTSizing. The best tested knowledge-injection configuration further reduces mean runtime by 39.9\% and samples by 35.4\%. This provides a route for advances in LLM circuit reasoning to benefit numerical sizing.

\section{Acknowledgment}
\label{sec:acknowledgment}

The authors used OpenAI GPT-6 Astra to assist with project code
development and language polishing of the manuscript. All
AI-assisted code and text were reviewed and revised as needed
by the authors, who take full responsibility for the final work.
All figures were drawn by the authors without generative AI.
The use of LLMs as circuit-knowledge advisers in the research
is described in Sections III-D and IV-D.

\end{document}